\documentclass[a4paper,11pt]{article}
\usepackage{pos}

\title{Measurement of electromagnetic transition form factors in two-photon collisions at BESIII}

\author*[a]{Max Lellmann}\onbehalf{on behalf of the BESIII collaboration}
\affiliation[a]{Institute for Nuclear Physics, Johannes Gutenberg-Universität Mainz}

\emailAdd{lellmann@uni-mainz.de}

\abstract{The muon anomalous magnetic moment, $a_\mu = (g-2)_\mu/2$, shows a sizeable discrepancy between Standard Model (SM) predictions and direct measurements, hinting at either New Physics or limitations in modeling low-energy strong interactions. As the uncertainty in the SM predictions currently exceeds the experimental uncertainty by a substantial margin, further improvements in theoretical precision are essential to clarify the origin of this discrepancy.

The SM theoretical uncertainty is dominated by hadronic processes. The hadronic light-by-light (HLbL) scattering contribution relies on highly accurate data for hadron production via two-photon fusion. Operating at $\sqrt{s}=3.773\,\text{GeV}$ with over $20\,\text{fb}^{-1}$ of integrated luminosity, the BESIII experiment is ideally suited to probe these processes at small momentum transfers. By precisely measuring channels like $e^+e^-\to e^+ e^-\pi^0$ and $e^+e^-\to e^+ e^-\pi^0\pi^0$, BESIII provides the critical experimental input necessary to sharpen the SM prediction and clarify the true nature of the $a_\mu$ anomaly.}

\FullConference{14th International Workshop on e$^+$e$^-$ collisions from Phi to Psi 2026 (PHIPSI26)\\
8-11 June 2026\\
Pisa, Italy\\}

\begin{document}
\maketitle

\section{Introduction}
Two-photon transition form factors (TFFs), $F(Q_1^2,Q_2^2)$, probe hadron structure and are critical for evaluating the hadronic light-by-light (HLbL) contribution to the muon anomalous magnetic moment, $a_\mu$ \cite{Aliberti:2025beg,Danilkin:2019mhd}. In order to probe new physics using $a_\mu$, the uncertainty of the SM prediction must be reduced. Using data-driven approaches on calculating the hadronic Light-by-Light scattering contribution, it is observed that pseudoscalar exchanges dominate the central value of $a_\mu^\text{HLbL}$, while, currently, light axial and tensor states drive its uncertainty. This quantities are not accessible using perturbative QCD at the important small energy scales. Consequently, experimental TFF measurements at low momentum transfer ($Q^2\lesssim1\,$GeV$^2$) are essential on improving the data-driven approach on calculating $a_\mu^\text{HLbL}$.
In the spacelike regime, TFFs are accessed via $e^+e^-\to e^+e^-X$ scattering \cite{HadroTOPS}, where the final-state lepton kinematics dictate the photon virtualities. Because perturbative QCD is inapplicable at small $Q^2$, the BESIII experiment \cite{BESIII:2009fln} --operating in the $\tau$-charm energy region—provides an ideal facility for studying hadronic two-photon interactions in the kinematic regime most relevant to $a_\mu$.
\section{\boldmath Measurement of the $\pi^0$ Transition Form Factor}
\begin{figure}[tb]
    \centering
    \includegraphics[width=.99\linewidth]{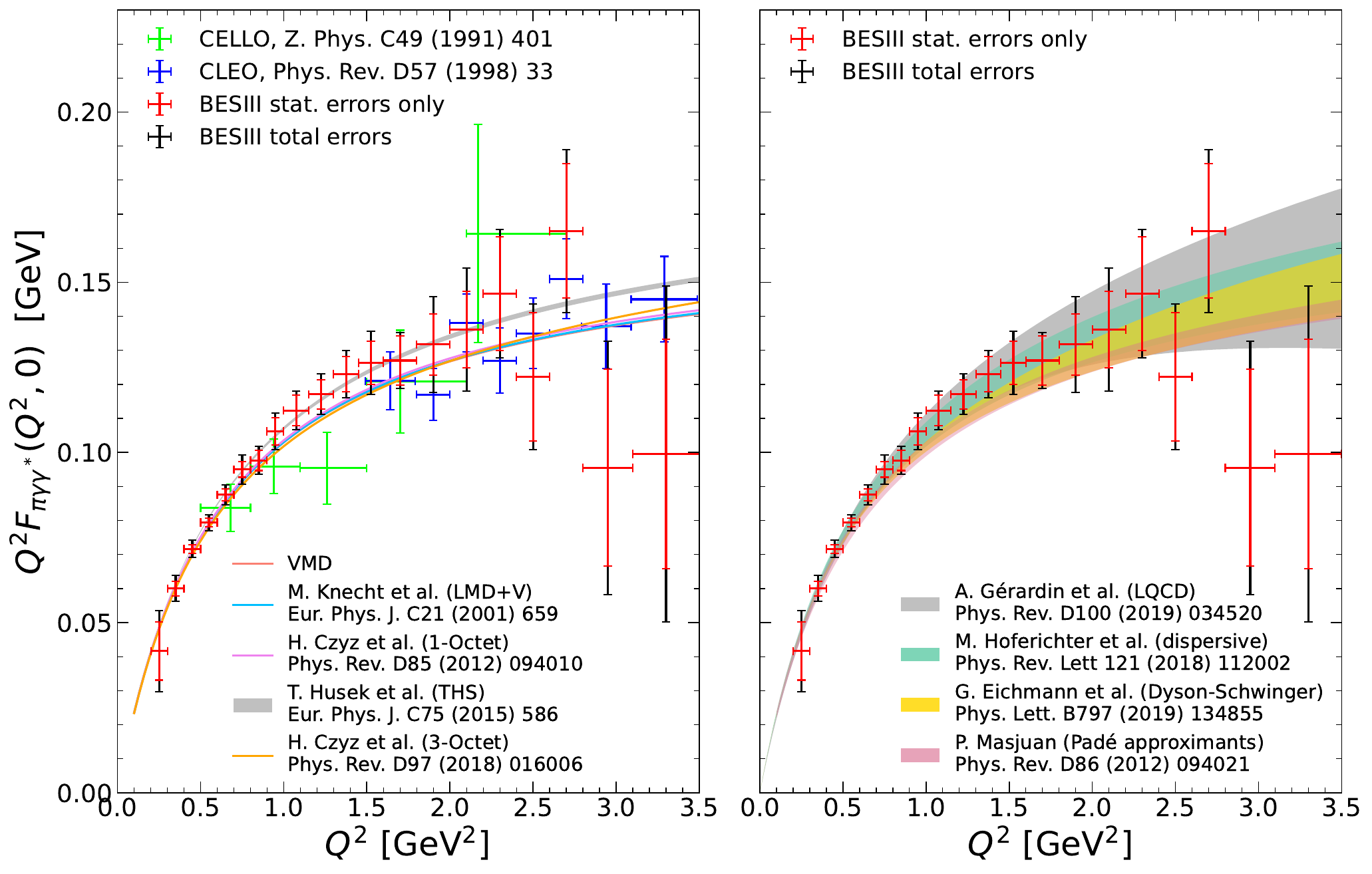}
    \caption{Result of the $\pi^0$ TFF measurement with statistical (red error bars) and total uncertainties (black error bars) compared
to previous measurements \cite{CELLO:1990klc,CLEO:1997fho} and hadronic models \cite{Knecht:2001xc,Czyz:2012nq,Husek:2015wta,Czyz:2017veo} (left panel) and predictions from dispersive \cite{Hoferichter:2018dmo}, lattice QCD \cite{Gerardin:2019vio}, and Dyson-Schwinger \cite{Eichmann:2019tjk} theory calculations.}
    \label{fig:pi0TFF}
\end{figure}
Using 2.93\,fb$^{-1}$ of $e^+e^-$ collision data at $\sqrt{s} = 3.773$\,GeV, the BESIII collaboration achieved the most precise measurement of the spacelike $\gamma\gamma^\ast\to\pi^0$ TFF at small $Q^2<2\,$GeV$^2$ (Fig.~\ref{fig:pi0TFF}) \cite{BESIII:2025zjx}. The analysis utilizes a single-tag technique for the $e^+e^-\to e^+e^-\pi^0$ process, detecting only the $\pi^0$ and one lepton. The untagged lepton escapes at small angles, constraining the virtuality of the corresponding photon to nearly zero. This enables a high-precision extraction of the single-virtual TFF, providing a stringent test for phenomenological and lattice QCD predictions. Future updates using a $20.3$\,fb$^{-1}$ dataset at the same energy will further enhance this precision and allows first measurements of the doubly-virtual TFF.

\section{\boldmath Study of Transitions Form Factors in the $\gamma\gamma^\ast\to\pi^0\pi^0$ Process}
\begin{figure}[tb]
    \centering
    \includegraphics[width=\linewidth]{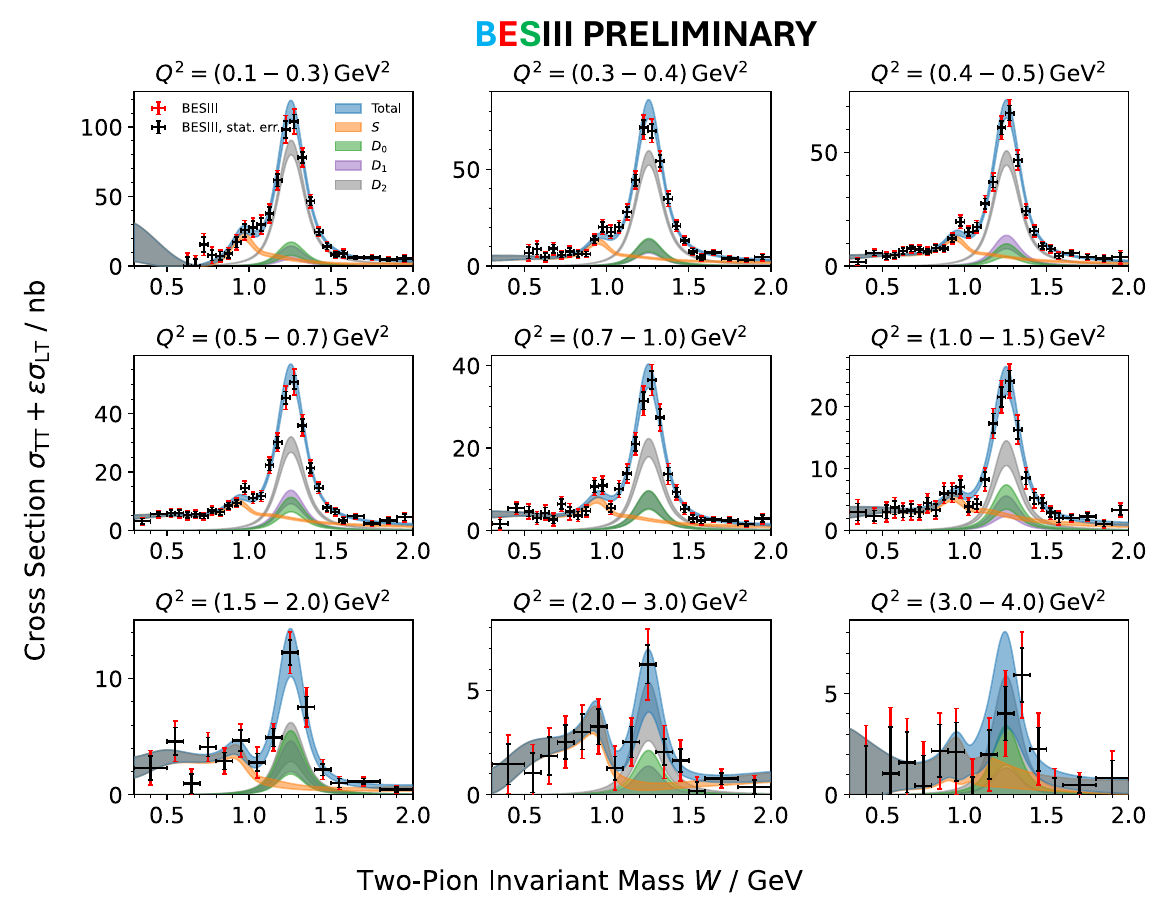}
    \caption{Comparison of the fit result (blue band) and the $S$ (orange), $D_0$ (green), $D_1$ (violet), and $D_2$ (gray) partial-wave contributions to the integrated singly virtual $\gamma\gamma^\ast\to\pi^0\pi^0$ BESIII data. Statistical (black) and total (red) uncertainties are shown separately. The widths of the shaded bands indicate the $1\sigma$ confidence intervals.}
    \label{fig:ST_pi0pi0}
\end{figure}
The TFFs of the scalar states $f_0(500)$ and $f_0(980)$ and the tensor meson $f_2(1270)$ can be studied in the process $\gamma\gamma^\ast\to\pi^0\pi^0$, where the resonances appear as intermediate states. At BESIII this space-like two-photon production of neutral pion pairs is analyzed using single-tagged $e^+e^-\to e^+e^-\pi^0\pi^0$ events from a $20.3$\,fb$^{-1}$ dataset at $\sqrt{s}=3.773\,$GeV. The $\gamma\gamma^\ast\to\pi^0\pi^0$ cross sections are extracted differentially in two-pion invariant mass $W$, tagged-lepton momentum transfer $Q^2$, and pion polar angle in the two-photon centre-of-mass system $\theta^\ast$. 

Because the off-shell photon can be longitudinally polarized, the measured cross section comprises a fully transverse term ($\sigma_\text{TT}$) and a mixed transverse-longitudinal term ($\sigma_\text{LT}$) \cite{HadroTOPS}. To account for their differing kinematic dependencies to the $e^+e^-\to e^+e^-\pi^0\pi^0$ cross section, a factor $\varepsilon$ is introduced, such that the combined observable $\sigma_\text{TT} + \varepsilon\sigma_\text{LT}$ is extracted. As shown in Fig.~\ref{fig:ST_pi0pi0}, this observable in $W$ and $Q^2$ clearly resolves the $f_2(1270)$ tensor state, alongside indications of the $f_0(500)$ and $f_0(980)$ scalars. Notably, this marks the first measurement of the $\gamma\gamma^\ast\to\pi^0\pi^0$ cross section for $Q^2<3\,$GeV$^2$.

\begin{figure}[tb]
    \centering
    \includegraphics[width=\linewidth]{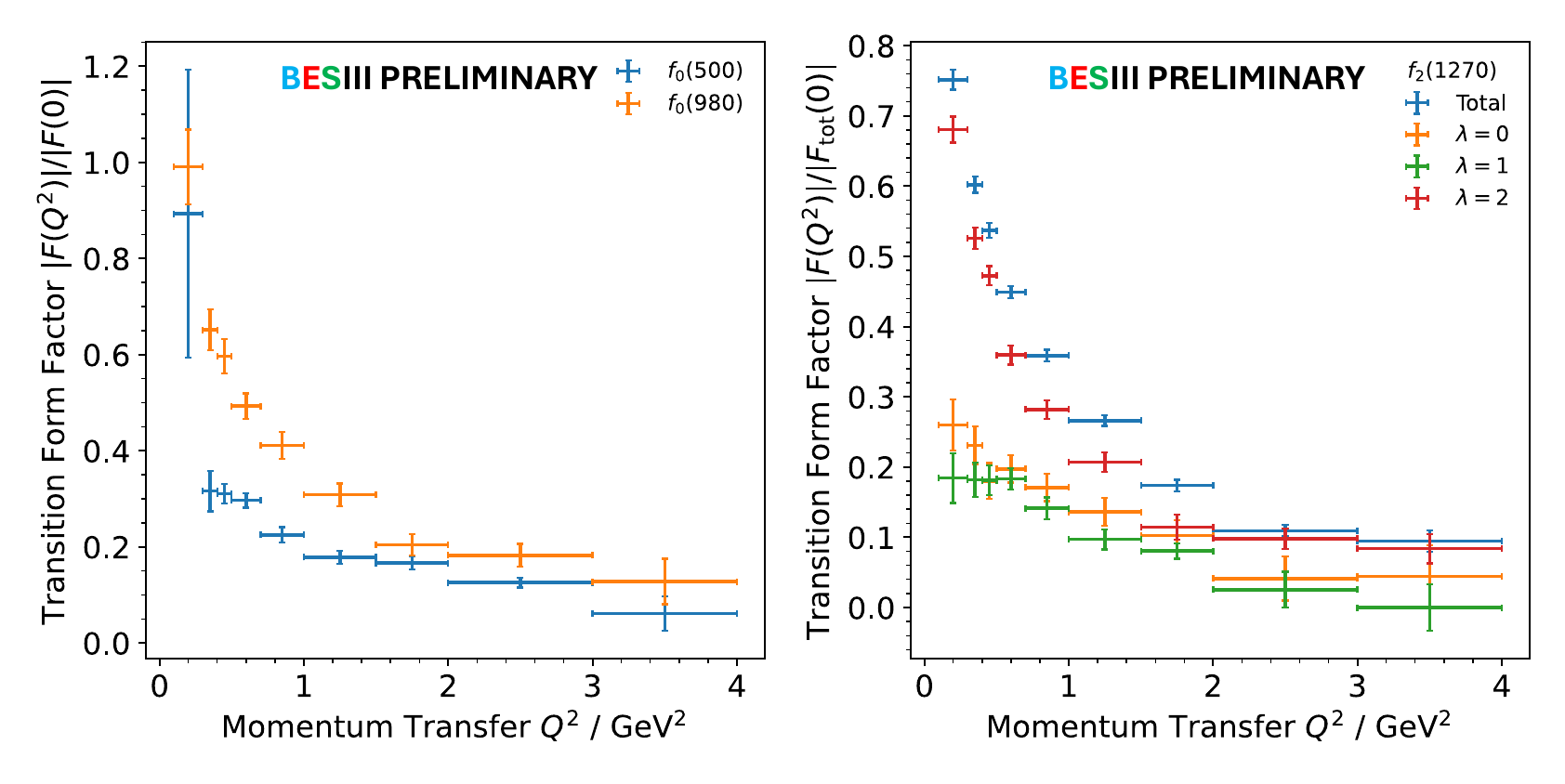}
    \caption{Transition form factors of $f_0(500)$ and $f_0(980)$ as a function of the momentum transfer $Q^2$, normalized to their values at $Q^2=0$ (left), and the total form factor of $f_2(1270)$ together with its helicity contributions as a function of the momentum transfer $Q^2$, normalized to the total form factor at $Q^2=0$ (right). The error bars show the total uncertainty.}
    \label{fig:TFF_pi0pi0}
\end{figure}

To extract the TFFs of resonances, a partial wave analysis is performed. The differential cross section $d(\sigma_\text{TT} + \varepsilon\sigma_\text{LT})/d|\cos\theta^\ast|$ is parameterized using spherical harmonics $Y_\ell^m(\theta^\ast)$:
$$d(\sigma_\text{TT} + \varepsilon\sigma_\text{LT})/d|\cos\theta^\ast| = 4\pi\left(1+Q^2/W^2\right)\Big(|S\,Y_0^0+D_0\,Y_2^0\,e^{i\phi}|^2+\varepsilon\,|D_1\,Y_2^1|^2+|D_2\,Y_2^2|^2\Big)\ ,$$
where the scalar amplitude $S$ is modeled via a coupled-channel Omn\`es formalism \cite{Danilkin:2019opj,Danilkin:2020pak} and the $f_2(1270)$ helicity amplitudes $D_{0,1,2}$ use a relativistic Breit-Wigner. The TFFs are defined by the amplitude ratio $A(Q^2)/A(0)$ at their respective pole positions, with the tensor helicity components constrained to $\sum_\lambda |F^\lambda_{f_2}|^2=|F_{f_2}^{\rm tot}|^2$. 
To fix the $Q^2=0$ normalization, this parameterization is fitted simultaneously to the single-tagged BESIII data and existing real-photon ($\gamma\gamma\to\pi^0\pi^0$) cross sections from BESIII (preliminary) \cite{Kussner:2024ryb}, Belle \cite{Belle:2009ylx}, and Crystal Ball \cite{CrystalBall:1990oiv}. The fit takes into account different detector resolutions of the two-pion invariant mass.
The extracted TFFs, shown in Fig.~\ref{fig:TFF_pi0pi0}), yield the first-ever measurement for the $f_0(500)$ TFF, alongside the most precise and lowest-$Q^2$ determinations for the $f_0(980)$ and $f_2(1270)$ TFFs. These latter results are fully consistent with previous Belle measurements at $Q^2=3-4\,$GeV$^2$ \cite{Belle:2015oin}.

\section{Conclusion and Outlook}
BESIII offers the ideal testbed for the measurement of transition form factors of light mesons at small to intermediate momentum transfers. By exploiting large data samples, the transition form factors of $\pi^0$, $\eta^\prime$ \cite{BESIII:2026bks}, $f_0(500)$, $f_0(980)$, and $f_2(1270)$ are determined at unprecedented precision.
The large data samples collected with BESIII, will not only enable double virtual determinations of the TFF (pseudo-)scalar and tensor states, but also the determination of the $f_1(1285)$ TFF, which is most critical for improving the Standard Model prediction of $a_\mu^\text{HLbL}$.

%
%
%
\bibliographystyle{JHEP}
\bibliography{skeleton-bib}

@article{Aliberti:2025beg,
    author = "Aliberti, R. and others",
    title = "{The anomalous magnetic moment of the muon in the Standard Model: an update}",
    eprint = "2505.21476",
    archivePrefix = "arXiv",
    primaryClass = "hep-ph",
    reportNumber = "CERN-TH-2025-101, FERMILAB-PUB-25-0344-T, INT-PUB-25-015, IPARCOS-UCM-25-029, KEK Preprint 2025-22, LTH 1403, MITP-25-037, UWThPh 2025-15, UWThPh
  2025-15, ZU-TH 37/25, IPARCOS-UCM-25-029",
    doi = "10.1016/j.physrep.2025.08.002",
    journal = "Phys. Rept.",
    volume = "1143",
    pages = "1--158",
    year = "2025"
}

@article{Danilkin:2019mhd,
    author = "Danilkin, Igor and Redmer, Christoph Florian and Vanderhaeghen, Marc",
    title = "{The hadronic light-by-light contribution to the muon{\textquoteright}s anomalous magnetic moment}",
    eprint = "1901.10346",
    archivePrefix = "arXiv",
    primaryClass = "hep-ph",
    doi = "10.1016/j.ppnp.2019.05.002",
    journal = "Prog. Part. Nucl. Phys.",
    volume = "107",
    pages = "20--68",
    year = "2019"
}

@article{HadroTOPS,
    author = "Lellmann, Max and Danilkin, Igor and Denig, Achim and Muskalla, Jan and Redmer, Christoph F. and Ren, Xiu-Lei and Vanderhaeghen, Marc",
    title = "{HadroTOPS: A Monte Carlo event generator for hadron production in two-photon scattering in electron positron collisions}",
    eprint = "2511.12717",
    archivePrefix = "arXiv",
    primaryClass = "hep-ph",
    doi = "10.1016/j.cpc.2026.110253",
    journal = "Comput. Phys. Commun.",
    volume = "327",
    pages = "110253",
    year = "2026"
}

@article{BESIII:2009fln,
    author = "Ablikim, M. and others",
    collaboration = "BESIII",
    title = "{Design and Construction of the BESIII Detector}",
    eprint = "0911.4960",
    archivePrefix = "arXiv",
    primaryClass = "physics.ins-det",
    doi = "10.1016/j.nima.2009.12.050",
    journal = "Nucl. Instrum. Meth. A",
    volume = "614",
    pages = "345--399",
    year = "2010"
}

@article{BESIII:2025zjx,
    author = "Ablikim, Medina and others",
    collaboration = "BESIII",
    title = "{Measurement of the space-like $\pi^0$ transition form factor}",
    eprint = "2509.07685",
    archivePrefix = "arXiv",
    primaryClass = "hep-ex",
    month = "9",
    year = "2025"
}

@article{CELLO:1990klc,
    author = "Behrend, H. J. and others",
    collaboration = "CELLO",
    title = "{A Measurement of the pi0, eta and eta-prime electromagnetic form-factors}",
    reportNumber = "DESY-90-110",
    doi = "10.1007/BF01549692",
    journal = "Z. Phys. C",
    volume = "49",
    pages = "401--410",
    year = "1991"
}

@article{CLEO:1997fho,
    author = "Gronberg, J. and others",
    collaboration = "CLEO",
    title = "{Measurements of the meson - photon transition form-factors of light pseudoscalar mesons at large momentum transfer}",
    eprint = "hep-ex/9707031",
    archivePrefix = "arXiv",
    reportNumber = "SLAC-PUB-9838, CLNS-97-1477, CLEO-97-7",
    doi = "10.1103/PhysRevD.57.33",
    journal = "Phys. Rev. D",
    volume = "57",
    pages = "33--54",
    year = "1998"
}

@article{Knecht:2001xc,
    author = "Knecht, Marc and Nyffeler, Andreas",
    title = "{Resonance estimates of O(p**6) low-energy constants and QCD short distance constraints}",
    eprint = "hep-ph/0106034",
    archivePrefix = "arXiv",
    reportNumber = "CPT-2001-P-4149",
    doi = "10.1007/s100520100755",
    journal = "Eur. Phys. J. C",
    volume = "21",
    pages = "659--678",
    year = "2001"
}

@article{Czyz:2012nq,
    author = "Czyz, Henryk and Ivashyn, Sergiy and Korchin, Alexandr and Shekhovtsova, Olga",
    title = "{Two-photon form factors of the $\pi^0$, $\eta$ and $\eta'$ mesons in the chiral theory with resonances}",
    eprint = "1202.1171",
    archivePrefix = "arXiv",
    primaryClass = "hep-ph",
    reportNumber = "IFIC-11-63, FTUV-2011-1110",
    doi = "10.1103/PhysRevD.85.094010",
    journal = "Phys. Rev. D",
    volume = "85",
    pages = "094010",
    year = "2012"
}

@article{Czyz:2017veo,
    author = "Czy{\.z}, Henryk and Kisza, Patrycja and Tracz, Szymon",
    title = "{Modeling interactions of photons with pseudoscalar and vector mesons}",
    eprint = "1711.00820",
    archivePrefix = "arXiv",
    primaryClass = "hep-ph",
    doi = "10.1103/PhysRevD.97.016006",
    journal = "Phys. Rev. D",
    volume = "97",
    number = "1",
    pages = "016006",
    year = "2018"
}

@article{Husek:2015wta,
    author = "Husek, Tomas and Leupold, Stefan",
    title = "{Two-hadron saturation for the pseudoscalar{\textendash}vector{\textendash}vector correlator and phenomenological applications}",
    eprint = "1507.00478",
    archivePrefix = "arXiv",
    primaryClass = "hep-ph",
    doi = "10.1140/epjc/s10052-015-3778-x",
    journal = "Eur. Phys. J. C",
    volume = "75",
    number = "12",
    pages = "586",
    year = "2015"
}

@article{Hoferichter:2018dmo,
    author = "Hoferichter, Martin and Hoid, Bai-Long and Kubis, Bastian and Leupold, Stefan and Schneider, Sebastian P.",
    title = "{Pion-pole contribution to hadronic light-by-light scattering in the anomalous magnetic moment of the muon}",
    eprint = "1805.01471",
    archivePrefix = "arXiv",
    primaryClass = "hep-ph",
    reportNumber = "INT-PUB-18-016",
    doi = "10.1103/PhysRevLett.121.112002",
    journal = "Phys. Rev. Lett.",
    volume = "121",
    number = "11",
    pages = "112002",
    year = "2018"
}

@article{Gerardin:2019vio,
    author = "G{\'e}rardin, Antoine and Meyer, Harvey B. and Nyffeler, Andreas",
    title = "{Lattice calculation of the pion transition form factor with $N_f=2+1$ Wilson quarks}",
    eprint = "1903.09471",
    archivePrefix = "arXiv",
    primaryClass = "hep-lat",
    reportNumber = "MITP-19-014",
    doi = "10.1103/PhysRevD.100.034520",
    journal = "Phys. Rev. D",
    volume = "100",
    number = "3",
    pages = "034520",
    year = "2019"
}

@article{Eichmann:2019tjk,
    author = "Eichmann, Gernot and Fischer, Christian S. and Weil, Esther and Williams, Richard",
    title = "{Single pseudoscalar meson pole and pion box contributions to the anomalous magnetic moment of the muon}",
    eprint = "1903.10844",
    archivePrefix = "arXiv",
    primaryClass = "hep-ph",
    doi = "10.1016/j.physletb.2019.134855",
    journal = "Phys. Lett. B",
    volume = "797",
    pages = "134855",
    year = "2019",
    note = "[Erratum: Phys.Lett.B 799, 135029 (2019)]"
}

@article{Danilkin:2019opj,
    author = "Danilkin, Igor and Deineka, Oleksandra and Vanderhaeghen, Marc",
    title = "{Dispersive analysis of the $\gamma^{*}\gamma^{*} \to \pi \pi$ process}",
    eprint = "1909.04158",
    archivePrefix = "arXiv",
    primaryClass = "hep-ph",
    doi = "10.1103/PhysRevD.101.054008",
    journal = "Phys. Rev. D",
    volume = "101",
    number = "5",
    pages = "054008",
    year = "2020"
}

@article{Danilkin:2020pak,
    author = "Danilkin, Igor and Deineka, Oleksandra and Vanderhaeghen, Marc",
    title = "{Data-driven dispersive analysis of the {\ensuremath{\pi}}{\ensuremath{\pi}} and {\ensuremath{\pi}}K scattering}",
    eprint = "2012.11636",
    archivePrefix = "arXiv",
    primaryClass = "hep-ph",
    doi = "10.1103/PhysRevD.103.114023",
    journal = "Phys. Rev. D",
    volume = "103",
    number = "11",
    pages = "114023",
    year = "2021"
}

@article{Kussner:2024ryb,
    author = {K{\"u}{\ss}ner, Meike},
    collaboration = "BESIII",
    title = "{Recent Highlights on Meson Spectroscopy at BESIII}",
    doi = "10.1051/epjconf/202429101002",
    journal = "EPJ Web Conf.",
    volume = "291",
    pages = "01002",
    year = "2024"
}

@article{Belle:2009ylx,
    author = "Uehara, S. and others",
    collaboration = "Belle",
    title = "{High-statistics study of neutral-pion pair production in two-photon collisions}",
    eprint = "0903.3697",
    archivePrefix = "arXiv",
    primaryClass = "hep-ex",
    reportNumber = "KEK-PREPRINT-2008-50, BELLE-PREPRINT-2009-4",
    doi = "10.1103/PhysRevD.79.052009",
    journal = "Phys. Rev. D",
    volume = "79",
    pages = "052009",
    year = "2009"
}

@article{CrystalBall:1990oiv,
    author = "Marsiske, H. and others",
    collaboration = "Crystal Ball",
    title = "{A Measurement of $\pi^0 \pi^0$ Production in Two Photon Collisions}",
    reportNumber = "SLAC-PUB-5163, DESY-90-002",
    doi = "10.1103/PhysRevD.41.3324",
    journal = "Phys. Rev. D",
    volume = "41",
    pages = "3324",
    year = "1990"
}

@article{Belle:2015oin,
    author = "Masuda, M. and others",
    collaboration = "Belle",
    title = "{Study of $\pi^0$ pair production in single-tag two-photon collisions}",
    eprint = "1508.06757",
    archivePrefix = "arXiv",
    primaryClass = "hep-ex",
    reportNumber = "BELLE-PREPRINT-2015-15, KEK-PREPRINT-2015-24",
    doi = "10.1103/PhysRevD.93.032003",
    journal = "Phys. Rev. D",
    volume = "93",
    number = "3",
    pages = "032003",
    year = "2016"
}

\end{document}